\documentclass[oneocolumn,showpacs,showkeys,preprintnumbers,amsmath,amssymb]{revtex4}

\usepackage{graphicx}
\usepackage{dcolumn}
\usepackage{bm}

\def\betabfeq{{\mbox{\boldmath $ \beta$}}}

\newcounter{saveeqn}

\begin{document}

\title{{ \bf  Reply to the comment on "Exact Expression for Radiation of an Accelerated Charge in Classical Electrodynamics"} }

\author{Young-Sea Huang}
\email{yshuang@mail.scu.edu.tw}
 \affiliation{Department of Physics, Soochow University, Shih-Lin, Taipei 111, Taiwan
}

\vspace{25mm}

\begin{abstract}
 Flaws and ambiguities are pointed out upon examining the comment attempting to solve a problem as raised recently --- the currently accepted formulation of electromagnetic radiation of an accelerated charge violates the principle of conservation of energy. This problem is not solved by the comment, due to  a misunderstanding in the meaning of the total radiated power crossing a sphere. An experiment is suggested to determine whether on not the currently accepted formulation is valid.

\end{abstract}

\pacs{
  41.60.-m radiation by moving charges;  03.30.+p special relativity.}

\keywords{electromagnetic radiation,  Larmor formula, Doppler effect,
  special relativity.}

\maketitle

  In a recent literature \cite{Singal1}, without any reason, Singal made a hasty conclusion that the newly derived exact expression for the electromagnetic radiation of an accelerated charge by Huang and  Lu \cite{Huang1} is incorrect. In the comment \cite{Singal2},  Singal attempted to resolve a serious problem --- the currently accepted formulation of electromagnetic radiation of an accelerated charge violates the principle of conservation of energy, as raised recently \cite{Huang2}. With a view to making a comprehensible reply, relevant expressions of both currently accepted formulation \cite{1,2,3,4,5}  and the new formulation \cite{Huang1}, as well as discrepancy between them are first presented. 

\section{\label{sec:1} The currently accepted formulation versus the new formulation}

According to the currently accepted formulation, the energy flux density of radiation is 
\begin{equation}\label{eqS}
 {\bf S} ({\bf r},t) 
= { q^{2}  \over 16 \, \pi^{2}\, \epsilon_{0}  \, c}   { (  \hat{ {\bf r}} \times (\, (\hat{ {\bf r} }-  {\betabfeq} ) \times \dot{\betabfeq}\,  )\, )^{2} \over  r^{2} \, (1- {\hat{\bf r}} \cdot \betabfeq )^{6} }  \,\, \hat{ {\bf r}}. 
\end{equation}
The energy flux density of radiation at time $t$ and at any position $ {\bf r}$ on a sphere, Eq.~(\ref{eqS}), is due to the radiation emanating by an accelerated charge at an earlier time $t_{r}=t-r/c$ and at the center of the sphere.  The velocity  $\betabfeq$ and the acceleration  $\dot{ \betabfeq}$ of the charge in the right-hand  side of Eq.~(\ref{eqS}) are defined  at the retarded time $t_{r}=t-r/c$. 
 The energy  radiated into a solid angle $d \Omega$ in the direction $\hat{ {\bf r}}$, and then measured at the position ${\bf r}$ and the time $t$ is $dW({\bf r},t) ={\bf S} \cdot {\hat {\bf r}} \,  r^{2}\, d \Omega \,dt$,  for an infinitesimal time interval $dt$. Hence, ${\bf S} \cdot {\hat {\bf r}} $ is the energy per unit area per unit time measured at the  position $ {\bf r}$ on the sphere. Consequently,  the radiated power passing through  the surface of a surrounding sphere per unit solid angle  $d \Omega$ in the direction $\hat{ {\bf r}}$ is
\begin{equation}\label{eqjak3}
 {d\, P({\hat {\bf r}},t) \over d\, \Omega}= \frac{ d\, W ({\bf r},t) }{ d\, \Omega \,\, d\, t } = {q^{2} \over 16 \, \pi^{2} \,\epsilon_{0} \, c}   { ( \, \hat{ {\bf r}} \times (\, (\hat{ {\bf r} }-  {\betabfeq} ) \times \dot{\betabfeq} \,  )\, )^{2} \over  (1- {\hat{\bf r}} \cdot \betabfeq )^{6} }. 
 \end{equation}
Eq.~(\ref{eqjak3}) is the angular distribution of  radiated power per unit solid angle, as measured by observers on a surrounding sphere \cite{5}.

 In contrast, according to the new formulation, the energy flux density of radiation is 
\begin{equation}\label{huangeq36}
\tilde{\bf S} ({\bf r},t) 
= { q^{2}  \over 16 \, \pi^{2}\, \epsilon_{0}  \, c}    {\gamma^{2} \, (  \hat{ {\bf r}} \times (\, (\hat{ {\bf r} }-  {\betabfeq} ) \times \dot{\betabfeq} \,  )\, )^{2} \over  r^{2} \, (1- {\hat{\bf r}} \cdot \betabfeq )^{4} }  \,\, \hat{ {\bf r}}. 
\end{equation}
Thus, the angular distribution of  radiated power per unit solid angle, relative to the position of the charge at the retarded time $t_{r}=t-r/c$, is
\begin{equation}\label{huangeqjak3}
 {d\, \tilde{P}({\hat {\bf r}},t) \over d\, \Omega}= { q^{2}  \over 16 \, \pi^{2}\, \epsilon_{0}  \, c}    {\gamma^{2} \, ( \, \hat{ {\bf r}} \times (\, (\hat{ {\bf r} }-  {\betabfeq} ) \times  \dot{\betabfeq} \,  )\, )^{2} \over  (1- {\hat{\bf r}} \cdot \betabfeq )^{4} }. 
\end{equation}
By  integrating  Eq.~(\ref{huangeqjak3})  over a surrounding sphere,  the total  radiated power crossing  any surrounding sphere is
\begin{equation}\label{huangeq38}
 \tilde{P} = { q^{2} \, \gamma^{6}  \over 6 \, \pi \,\epsilon_{0}\, c}   (  \dot{\betabfeq}^{2} - ( {\betabfeq} \times  \dot{\betabfeq})^{2} \,) . 
\end{equation}
The total radiated  power of radiation crossing  any surrounding sphere is equal to the total power emitted by the charge at the retarded time $t_{r}$, the so-called Li\'{e}nard's result,
\begin{equation}\label{eq38s}
P(t_{r}) = { q^{2} \, \gamma^{6}  \over 6 \, \pi \,\epsilon_{0}\, c}   (  \dot{\betabfeq}^{2} - ( {\betabfeq} \times  \dot{\betabfeq} )^{2} \,) , 
\end{equation}
as it should be to fulfill the principle of conservation of energy.
 Yet, according to the currently accepted formulation, by integrating  Eq.~(\ref{eqjak3})  over a surrounding sphere,  the total power of radiation crossing any surrounding sphere is not equal to the Li\'{e}nard's result \cite{Huang2}. The currently accepted formulation violates the principle of conservation of energy.

\section{\label{sec:2}  Singal's resolution of the problem --- the currently accepted formulation violates the principle of conservation of energy}
 
\begin{figure}[ht]
\begin{center}
\includegraphics[width=0.45\textwidth, clip=]{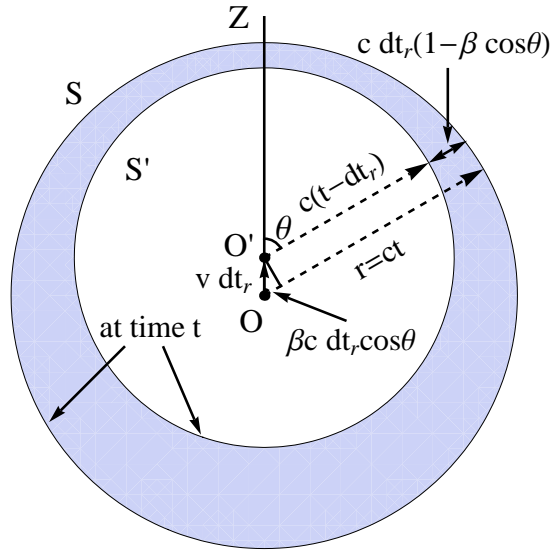}
\end{center}
\caption{\label{fig1}
 Suppose at time $t=0$, an accelerated charge is at the origin $O$ and moves with a velocity $v$ along the z-axis. In a time interval $dt_{r}$, the change moves a distance $v\, dt_{r}$ to  position $O'$ in the z-axis. At time $t$, the radiation in the region (shade area) enclosed by spheres $S'$ and $S$ is the radiation emitted by the charge during the time interval from $0$ to $dt_{r}$. }
\end{figure}

In the comment, Singal first claimed that  our reasoning above is fallacious, because we equate the evaluated result of the total radiated power of radiation crossing  a sphere to the total power emitted by the charge. Then, he presented a resolution to the problem.  Referring to Fig.~\ref{fig1}, at time $t$ the radiation in the region enclosed by spheres $S'$ and $S$ is due to the radiation emitted by the charge during the time interval from $0$ to $dt_{r}$. The radiation region is not spherically symmetric around $O$, since the charge moves a distance $v\, dt_{r}$ from $O$ to $O'$. Thus, the radiation
emitted by the charge during the time interval $dt_{r}$ does not cross the sphere $S$ at all points in the time interval $dt_{r}$, rather it takes $dt=dt_{r}\, (1- {\hat{\bf r}} \cdot \betabfeq )$. 

Therefore, from Eq.~(\ref{eqjak3}),  the total energy of radiation ${\cal W}$ enclosed by spheres $S'$ and $S$ is evaluated as
\begin{equation}\label{Toteqjak3}
 \begin{array}{rcl}
 {\cal W}  &=& \int \frac{ d\, P (\hat{\bf r},t) }{ d\, \Omega  } d\, t_{r}(1- {\hat{\bf r}} \cdot \betabfeq )\, d\, \Omega \\  &=& d\, t_{r} \times  \int {q^{2} \over 16 \, \pi^{2} \,\epsilon_{0} \, c}   { ( \, \hat{ {\bf r}} \times (\, (\hat{ {\bf r} }-  {\betabfeq} ) \times \dot{\betabfeq} \,  )\, )^{2} \over  (1- {\hat{\bf r}} \cdot \betabfeq )^{5} }d\, \Omega \\
 &=& d\, t_{r} \times { q^{2} \, \gamma^{6}  \over 6 \, \pi \,\epsilon_{0}\, c}   (  \dot{\betabfeq}^{2} - ( {\betabfeq} \times  \dot{\betabfeq})^{2} \,). 
 \end{array} 
 \end{equation}
The total energy of radiation ${\cal W}$ is  due to the total  energy emitted by the charge during the time interval $dt_{r}$.  ${\cal W} / d\, t_{r}$ is equal to  the total power emitted by the charge at the retarded time $t_{r}$, that is,
\begin{equation}\label{Leinard}
 \frac{{\cal W} }{d\, t_{r}} =  P(t_{r}) = { q^{2} \, \gamma^{6}  \over 6 \, \pi \,\epsilon_{0}\, c}   (  \dot{\betabfeq}^{2} - ( {\betabfeq} \times  \dot{\betabfeq})^{2} \,). 
\end{equation}
Therefore, the currently accepted formulation does not violate the principle of conservation of energy. In that case, the new  formulation violates the principle of conservation of energy. 
Yet, it should be noted that ${\cal W} / d\, t_{r} $ is not the total radiated  power crossing the sphere $S$ at time $t$. According to Singal's reasoning, what is the total radiated  power crossing the sphere $S$?

\section{\label{sec:3} The problem of the currently accepted formulation remains unsolved}

In our reasoning \cite{Huang1,Huang2}, we do not derive the Li\'{e}nard's result, and not equate the evaluated result of the total radiated power crossing a sphere to the Li\'{e}nard's result.  Instead, we compare the evaluated  result with the Li\'{e}nard's result to see which formulation satisfies the principle of conservation of energy. 
It turns out that currently accepted formulation, instead of the new formulation, violates the principle of conservation of energy.

There are problems in the currently accepted formulation. Since the charge is accelerated, it does not move uniformly during the time interval $dt_{r}$. Thus, the angular distribution of radiated power in the time interval is not exactly in accordance with Eq.~(\ref{eqjak3}) as evaluated at merely one retarded time $t_{r}$, because the time-retarded positions and velocities $\betabfeq$ of the charge change during the time  interval. The total  power emitted by the charge as evaluated in accordance with Singal's reasoning should be only  approximately valid, as noticed by Panofsky and Jackson \cite{1,2}.
Therefore, it is very unlikely that the {\it exact} Li\'{e}nard's result is obtained simply  by  an  approximation approach, unless Eq.~(\ref{eqjak3}) is only approximately correct.  
 
According to Singal, the factor $(1- {\hat{\bf r}} \cdot \betabfeq)$ in the time interval $dt=dt_{r}\, (1- {\hat{\bf r}} \cdot \betabfeq)$ is interpreted as just a matter of simple geometry. Yet this interpretation negates the existent interpretations: a Lorentz transformation of time between the charge's frame and the observer's frame  \cite{2}, or something similar to the Doppler effect \cite{3}. Which interpretation is correct? 
 
   Foremost, Singal does not resolve the problem that the currently accepted formulation violates the principle of conservation of energy. The total radiated  power  crossing  a surrounding sphere at time $t$ is not equal to ${\cal W} / d\, t_{r} $. 
 Suppose that one wants to evaluate the total radiated {\it power} $P$ crossing a sphere. One first measures the total radiated  {\it energy} $W$ crossing the sphere in a time interval $\Delta t \,\, (\Delta t << 1)$.  Then, the total radiated power  crossing the sphere is given as $P=W / \Delta t$. It should be emphasized that the time interval $\Delta t$ must be  the same, as the measurement of the total radiated   energy  is carried out {\it at all points} on the sphere.  However, according to Singal's reasoning, the measurement of the total radiated energy $\cal{W}$ crossing the sphere $S$ is carried out with different time intervals $dt=dt_{r}\, (1- {\hat{\bf r}} \cdot \betabfeq)$ at different points on the sphere. ${\cal W} / dt$ is meaningless, and ${\cal W} / dt_{r}$ is not the total radiated power crossing the sphere. If Singal thinks that the total radiated power crossing  a surrounding sphere is ${\cal W} / d\, t_{r} $, and it is equal to the total power emitted by the charge at the retarded time, then he might make a mistake in the meaning of the total radiated power  crossing a sphere. Hence, the problem that the currently accepted formulation violates the principle of conservation of energy remains unsolved.

\section{\label{sec:4} The issue of the relativistic transformation of electromagnetic fields}
 
   Another issue in the comment is "While deriving expressions for the electric and magnetic fields, ${\bf E}$ and ${\bf B}$, Huang and Lu  \cite{Huang1} in their Eqs.~(20)-(25) simply replaced $r'$ with $r$ which is not correct as these two quantities are actually related by $r' = r / \delta$, where $\delta = 1/ \gamma (1- {\hat{\bf r}} \cdot \betabfeq)$ is the Doppler factor [3]. Thus their transformed electric and magnetic fields are wrong." The newly derived expressions for the electromagnetic fields become the currently accepted expressions, if  $r' = r \, \gamma\, (1- {\hat{\bf r}} \cdot \betabfeq)$, instead of $r' = r$, is employed in the transformation of electric and magnetic fields between inertial frames.

First, that the new formulation, rather than the currently accepted formulation, fulfills the principle of conservation of energy reinforces the validity of the replacement $r'=r$  in the transformation. Furthermore, Maxwell's equations of electrodynamics are shown form-invariant via a novel perspective on relativistic transformation --- transformation of physical quantities, instead of space-time coordinates \cite{YSHuang3}. An extra transformation of spacial coordinate such as $r' = r \, \gamma\, (1- {\hat{\bf r}} \cdot \betabfeq)$ is not necessary in the transformation of  electric and magnetic fields to render Maxwell's equations form-invariant among inertial frames.

 According to Singal, the expression $r' = r \, \gamma\, (1- {\hat{\bf r}} \cdot \betabfeq)$ is considered as due to the Doppler effect. Yet, the Doppler effect is  the transformation of physical quantities of waves such as  frequency $\nu$ and  wave vector ${\bf k}$ relative to inertial frames, instead of space-time coordinates $(t, {\bf r})$. The expression $r' = r \, \gamma\, (1- {\hat{\bf r}} \cdot \betabfeq)$ has nothing to do with the Doppler effect, since it involves  spatial coordinates only, without frequency  and  wave vector of waves.  A systematic method to derive the Doppler effect, without involving transformation of space-time coordinates, was presented in the literature \cite{Huang1,YSHuang5}. Even more, in a certain case an anomaly --- the problem of negative frequency of waves, was found by applying the invariance of the phase of waves which is equivalent to  relativistic transformation of both physical quantities $(\nu, {\bf k})$ and space-time coordinates  $(t, {\bf r})$ simultaneously \cite{YSHuang2}. This indicates that the invariance of the phase of waves is  invalid. Therefore, the Doppler effect should be related to the transformation of  physical quantities $(\nu, {\bf k})$ of waves only. In Singal's interpretation, the factor $(1- {\hat{\bf r}} \cdot \betabfeq)$ in $dt=dt_{r}\, (1- {\hat{\bf r}} \cdot \betabfeq)$ is  a matter of simple geometry, whereas  the factor $ \gamma\, (1- {\hat{\bf r}} \cdot \betabfeq)$ in $r' = r \, \gamma\, (1- {\hat{\bf r}} \cdot \betabfeq)$ is due to the Doppler effect. The two interpretations seem incompatible.

\section{\label{sec:5} Conclusion}

Ambiguities on the meaning of $dt=dt_{r}\, (1- {\hat{\bf r}} \cdot \betabfeq )$ and $r' = r \, \gamma\, (1- {\hat{\bf r}} \cdot \betabfeq)$ still exist in the currently accepted formulation of electromagnetic radiation of an accelerated charge. Owing to a misunderstanding in the meaning of the total radiated power crossing a sphere, the serious problem that currently accepted formulation violates the principle of conservation of energy remains unsolved. Such controversies as paradoxes in special relativity  are hardly resolved just by theoretical reasoning.  In order to convincingly determine the validity of the currently accepted formulation,  an experimental test on  the angular distribution of radiated power was proposed \cite{Huang2}. Nonetheless, it is necessary to clarify  which  the angular distribution of  radiated power is: Eq.~(\ref{eqjak3}), or from  Eq.~(\ref{Toteqjak3})
\begin{equation}\label{eqjak3a}
 {d\, P({\hat {\bf r}},t) \over d\, \Omega} = {q^{2} \over 16 \, \pi^{2} \,\epsilon_{0} \, c}   { ( \, \hat{ {\bf r}} \times (\, (\hat{ {\bf r} }-  {\betabfeq} ) \times \dot{\betabfeq} \,  )\, )^{2} \over  (1- {\hat{\bf r}} \cdot \betabfeq )^{5} }. 
 \end{equation} Further theoretical and experimental  examinations  on  the currently accepted formulation and the new formulation should be highly welcome.

\begin{acknowledgments}
We gratefully acknowledge Dr.  C.M.L.  Leonard  for valuable comments
during the preparation of this paper.

\end{acknowledgments}


\begin{thebibliography}{99}
\bibitem{Singal1} A.K. Singal,  "A first principles derivation of the electromagnetic fields of a point charge in arbitrary motion",  Am. J. Phys. {\bf 79}, 1036-1041 (2011).
\bibitem{Huang1} Young-Sea Huang and Kang-Hao Lu, "Exact Expression for Radiation of an Accelerated Charge in Classical Electrodynamics", Found. Phys. {\bf 38}, 151-159 (2008).
\bibitem{Singal2} A.K. Singal,  " Comment on "Exact Expression for Radiation of an Accelerated Charge in Classical Electrodynamics" ",  Found. Phys. {\bf 43}, 267-270 (2013).
\bibitem{Huang2} Young-Sea Huang, "Is the current formulation of the Radiation of an Accelerated Charge valid? ", Nuovo Cimento B {\bf 124}, 925-929 (2009).
\bibitem{1}  W.K.H. Panofsky and M. Phillips, {\it Classical Electricity and Magnetism}, (Addison-Wesley Publishing, Inc., 1962), Chapter 20.
\bibitem{2}  J.D. Jackson, {\it Classical Electrodynamics}, third edition, (John Wiley \& Sons Inc., New York, 1999), Chapter 14.
\bibitem{3} D.J. Griffiths, {\it Introduction to Electrodynamics}, 2nd edition, (Prentice-Hall, Inc., New Jersey, 1986), Chapters 9 and 10.
\bibitem{4} L.D. Landau, and E.M. Lifshitz, {\it The Classical Theory of Field}, fourth edition, (Pergamon Press Ltd., 1975), Chapter 9.
\bibitem{5} G.S. Smith, {\it An introduction to classical electromagnetic radiation}, (Cambridge University Press, New York, 1997), Chapter 6.
\bibitem{YSHuang3} Young-Sea Huang,  "A new perspective of relativistic transformation for Maxwell's equations of  electrodynamics", Phys. Scr.  {\bf 79}, 055001 (2009); "A new perspective on relativistic transformation versus the conventional Lorentz transformation illustrated by the problem of electromagnetic waves in moving media", {\bf 81}, 015004 (2010); "A new perspective on relativistic transformation: formulation of the differential Lorentz transformation based on first principles", {\bf 82}, 045011 (2010).
\bibitem{YSHuang5} Young-Sea Huang,  Can. J. Phys., "Formulation of the classical and the relativistic Doppler effect by a systematic method",  {\bf 82}, 957-964 (2004).
\bibitem{YSHuang2} Young-Sea Huang, "The invariance of the phase of waves among inertial frames is questionable", EPL  {\bf 79}, 10006 (2007); "Is the phase of plane waves an invariant?", Z. Naturforsch. {\bf 65a}, 615 (2010).


\end{thebibliography}
\end{document}